\documentclass[preprintnumbers,amsmath,amssymb,floatfix,12pt,superscriptaddress,nofootinbib]{revtex4}

\begin{document}
\title{New holographic Chaplygin gas model of dark energy}

\author{M. Malekjani\footnote{Email:\text{malekjani@basu.ac.ir}}}
\author{A.Khodam-Mohammadi\footnote{Email:\text{khodam@basu.ac.ir}}}
\affiliation{Department of Physics, Faculty of Science, Bu-Ali Sina
University, Hamedan 65178, Iran}

\begin{abstract}
\vspace*{1.5cm} \centerline{\bf Abstract} \vspace*{1cm} In this
work, we investigate the holographic dark energy model with new
infrared cut-off (new HDE model) proposed by Granda and Oliveros.
Using this new definition for infrared cut-off, we establish the
correspondence between new HDE model and standard Chaplygin gas
(SCG), generalized Chaplygin gas (GCG) and modified Chaplygin gas
(MCG) scalar field models in non-flat universe. The potential and
dynamics for these scalar field models, which describe the
accelerated expansion of the universe are reconstructed. According
to the evolutionary behavior of new HDE model, we derive the same
form of dynamics and potential for different SCG, GCG and MCG
models. We also calculate the squared sound speed of new HDE model
as well as for SCG, GCG and MCG models and investigate the new HDE
Chaplygin gas models from the viewpoint of linear perturbation
theory. All results in non-flat universe are also discussed in the
limiting case of flat universe, i.e. $k=0$.
\end{abstract}

\maketitle

\newpage
\section{Introduction}
Recent astronomical data from distant Ia supernova, Large Scale
Structure (LSS) and Cosmic Microwave Background (CMB) \cite{SN}
indicate that the current universe is not only expanding, but also
is experiencing an accelerated expansion. The accelerated expansion
can be driven by an exotic fluid with negative pressure, the
so-called dark energy (DE) \cite{copel, fri}. The nature of DE is
still unknown and many theoretical models have been suggested to
describe its behavior. Although, the simplest theoretical candidate
of DE is the cosmological constant with the equation of state
independent of cosmic time, $w=-1$ \cite{sahn}, but it suffers the
two well-known problems namely "fine-tuning" and "cosmic
coincidence" \cite{copel}. Both of these problems are related to the
DE density. In order to alleviate or even solve these problems, many
dynamical DE models have been suggested, whose equation of state and
their DE density are time-varying. It is worth noting that the
predictions of cosmological constant model is still fitted to the
current observation \cite{LCDM1}. Therefore, a suggested dynamical
DE model should not be faraway from the cosmological constant model.
Dynamical DE models can be classified into two categories i) The
scalar field DE models including quintessence \cite{Wetterich},
K-essence \cite{Chiba}, phantoms \cite{Caldwell1}, tachyon
\cite{Sen}, dilaton \cite{Gasperini}, quintom \cite{Elizalde1} and
so forth. ii) The interacting DE models, by considering the
interaction between dark matter and DE, including Chaplygin gas
\cite{Kamenshchik}, braneworld models \cite{Deffayet}, holographic
DE \cite{holo1} and agegraphic DE \cite{age1} models ,
etc.\\
The problem of DE has been investigated in the framework of string
theory and quantum gravity. Despite the lack of a complete theory in
quantum gravity, we can make some efforts to investigate the nature
of DE according to the principles of quantum gravity. The
holographic DE (HDE) and agegraphic DE (ADE) models have been
suggested based on the principle of quantum gravity theory. The ADE
model is based on the uncertainty relation of quantum mechanics
together with the gravitational effect in general relativity. The
HDE model is constructed based on the holographic principle
\cite{Cohen,Horava,holo1}. Based on the the validity of the
effective local quantum field theory, a short distance (UV) cut-off
$\Lambda $ is related to the long distance (IR) cut-off $L$ due to
the limit set by the formation of a black hole \cite{Cohen}.
Applying the UV-IR relationship constrains the total vacuum energy
in a box of volume $L^3$, where the total vacuum energy can not be
greater than the mass of a black hole with the same size. This upper
limit for vacuum energy density is given by
\begin{equation}
\rho_{\Lambda}\leq M_p^2L^{-2},
\end{equation}
where $M_p$ is the reduced Planck mass and $L$ is an IR cutoff.
Saturating the inequality, we obtain the HDE density
\begin{equation}
\rho_{\Lambda}=3c^2M_p^2L^{-2},
\end{equation}
where $c$ is a free dimensionless constant and the numeric
coefficient is chosen for convenience. The HDE model has been
constrained by various astronomical observation
\cite{obs3a,obs2,obs1,Wu:2007fs,obs3} and also investigated widely
in the literature \cite{nonflat,holoext,intde}. The IR cutoff $L$ is
related to the large scale of the universe, for example Hubble
horizon, future event horizon or particle horizon. If we take the
size of Hubble horizon or particle horizon as a length scale $L$ ,
the accelerated expansion of the universe can not be derived by HDE
model \cite{holo1}. However, in the case of event horizon, HDE model
can derive the universe with accelerated expansion \cite{holo1}. The
arising problem with the event horizon is that it is a global
concept of spacetime and existence of it depends on the future
evolution of the universe only for a universe with forever
accelerated expansion. Furthermore, the HDE with the event horizon
as a length scale is not compatible with the age of some old high
redshift objects \cite{zhang4}.\\Granda and Oliveros (GO, here
after) proposed a new IR cut-off containing the local quantities of
Hubble and time derivative Hubble scales \cite{garanda}. The
advantages of HDE with GO cutoff (new HDE model) is that it depends
on local quantities and avoids the causality problem appearing with
event horizon IR cutoff. The new HDE model can also obtain the
accelerated expansion of the universe \cite{garanda}. GO showed that
the transition redshift from deceleration phase ($q>0$) to
acceleration phase ($q<0$) is consistent with current observational
data \cite{garanda,observ}. The new HDE model has been extended into
the scalar field models both in flat and non-flat universe
\cite{granda2}. The correspondence between this model and scalar
fields allows us to reconstruct the potentials and the dynamics of
scalar fields. The new HDE model has also been investigated in
non-flat universe \cite{karami1}.\\
However, the early inflation era leads to a flat universe, but the
effect of curvature can not be neglected at present time.
Observationally, the CMB experiments preferred a closed universe
with small positive curvature \cite{Sie}. The WMAP analysis also
provides further confidence to show that a closed universe with
positively curved space is marginally preferred \cite{Uzan}.
Therefore, we are motivated to
investigate the new HDE model in a non flat universe.\\
On the other hand, the Chaplygin gas is one of the candidate of DE
models to explain the accelerated expansion of the universe. The
striking features of Chaplygin gas DE is that it can be assumed as a
possible unification of dark matter and DE. The Chaplygin gas plays
a dual role at different epoch of the history of the universe: it
can be as a dust-like matter in the early time, and as a
cosmological constant at late times. This model from the field
theory points of view is investigated in \cite{Bil}. The Chaplygin
gas emerges as an effective fluid associated with D-branes
\cite{Bor} and can also be obtained from the Born-Infeld action
\cite{Ben}. The simplest form of Chaplygin gas model called standard
Chaplygin gas (SCG) which has been used to explain the accelerated
expansion of universe \cite{gorii}. Although the SCG model can
interpret the accelerated expansion of universe, but it can not
explain the astrophysical problems such as structure formation and
cosmological perturbation power spectrum \cite{sanvik}.
Subsequently, the SCG is extended into the generalized Chaplygin gas
(GCG) which can construct viable cosmological models. The GCG model
is also modified into the modified Chaplygin gas (MCG) which can
show the radiation era in the early universe \cite{late}. The
correspondence between the HDE and ADE models
 with the Chaplygin gas energy density has been established in \cite{setar}.\\
  It should be noted that the correspondence between HDE and ADE
with SCG is problematic. This problem arises from the viewpoint of
liner perturbation theory. In this theory, the SCG model is stable
against small perturbation \cite{gori}, While the HDE and ADE have
an instability of a given perturbation.\cite{myung}. The crucial
quantity which shows the stability or instability of density
perturbation is squared speed of sound. For positive value of this
quantity, we have a regular propagating mode (stability), while
negative value of squared speed
shows an exponentially growing mode (instability) for a density perturbation.\\
 Here, in this work, we establish the connection
between the SCG, GCG and MCG models with new HDE model in non-flat
universe. This connection allows us to reconstruct the potentials
and the dynamics of the scalar fields according to evolutionary form
of new HDE to describe the SCG, GCG and MCG cosmology. We also
calculate the squared speed for new HDE model and discuss the
holographic interpretation of SCG, GCG and MCG models from the
viewpoint of linear perturbation theory. The results are also
discussed in the limiting case of flat universe.

\section{New HDE model}
The energy density of new HDE model is \cite{garanda}
\begin{equation}
\rho_{\Lambda}=3M_{P}^{2}(\alpha H^{2}+\beta\dot{H}),\label{rhoh}
\end{equation}
where $\alpha$ and $\beta$ are constants, $H$ is the Hubble
parameter and dot denotes the time derivative with respect to the
cosmic time. The Friedmann-Robertson-Walker (FRW) metric for a
universe with spatial curvature $k$ is
\begin{eqnarray}
 ds^2=dt^2-a^2(t)\left(\frac{dr^2}{1-kr^2}+r^2d\Omega^2\right),\label{metric}
 \end{eqnarray}
where $a(t)$ is the scale factor, and $k = -1, 0, 1$ represents the
open, flat, and closed universes, respectively. A closed universe
with small positive curvature ($\Omega_k\sim 0.02$) is compatible
with observation \cite{Bennett}. Like \cite{granda2}, we restrict
ourselves to the current DE dominated universe. Hence, the Freidmann
equation is written as
\begin{eqnarray}\label{Fried}
H^2+\frac{k}{a^2}=\frac{1}{3M_p^2}\rho_{\Lambda}.
\end{eqnarray}
Substituting Eq.(\ref{rhoh}) in (\ref{Fried}) yields
\begin{equation}\label{HDE1}
\frac{{\rm d}H^2}{{\rm
d}x}+\frac{2}{\beta}(\alpha-1)H^2=\frac{2}{\beta}ke^{-2x},
\end{equation}
where $x=\ln{a}$. Integrating Eq.(\ref{HDE1}) with respect to $x$
gives the following relation for Hubble parameter in new HDE
dominated non-flat universe
\begin{equation}
H^{2}=\frac{k}{\alpha-\beta-1} e^{-2x} +\gamma
e^{-\frac{2}{\beta}(\alpha-1)x},\label{H2-1}
\end{equation}
where $\gamma$ is an integration constant. From conservation
equation
\begin{equation}\label{contin}
\dot{\rho_{\Lambda}}+3H(1+\omega_{\Lambda})\rho_{\Lambda}=0,
\end{equation}
and using Eq.(\ref{rhoh}), one can easily obtain the equation of
state (EoS) parameter,
$\omega_{\Lambda}=p_{\Lambda}/\rho_{\Lambda}$, as
\begin{equation}
\omega_{\Lambda}=-1-\frac{2\alpha H \dot{H}+\beta\ddot{H}}{3H(\alpha
H^{2}+\beta \dot{H})}\label{wh1}
\end{equation}
Inserting $H$ from Eq.(\ref{H2-1}) in (\ref{wh1}) obtains
\begin{equation}
\omega_{\Lambda}=-\frac{1}{3}\left(\frac{k\Big(\frac{\alpha-\beta}{\alpha-\beta-1}\Big)e^{-2x}+\gamma\Big(\frac{3\beta-2\alpha+2}{\beta}\Big)e^{-\frac{2}{\beta}(\alpha-1)x}}
{k\Big(\frac{\alpha-\beta}{\alpha-\beta-1}\Big)e^{-2x}+\gamma
e^{-\frac{2}{\beta}(\alpha-1)x}}\right),\label{wh2}
\end{equation}
which expresses the time-dependent EoS parameter of new HDE in
non-flat universe. The time dependence of EoS parameter of DE allows it to transit
from $w_{\Lambda}>-1$ to $w_{\Lambda}<-1$ \cite{wang3}. The analysis
of DE models from the observational point of view indicates that a model with $w_{\Lambda}$
crossing $-1$ in the near past is favored \cite{alam2}.\\
Putting $k=0$ in Eqs. (\ref{H2-1}) and (\ref{wh2}), the Hubble
parameter and EoS parameter of new HDE model in flat universe are
reduced as
\begin{equation}
H^{2}=\gamma e^{-\frac{2}{\beta}(\alpha-1)x}\label{H2-flat}
\end{equation}
\begin{equation}
\omega_{\Lambda}=-1+\frac{2}{3}\frac{\alpha-1}{\beta},\label{wflat}
\end{equation}
Contrary to the non flat universe, the EoS parameter in flat
universe is constant with cosmic time. In order to obtain the
accelerated expansion of universe, $-1<w_{\Lambda}<-1/3$, the
constants $\alpha$ and $\beta$ must be limited as: $\beta>\alpha-1$
if $\alpha>1$ or $\beta<\alpha-1$ if $\alpha<1$. Also, the new HDE
can achieve the phantom phase, $w_{\Lambda}<-1$, for $\alpha<1$,
$\beta>0$ or $\alpha>1$, $\beta<0$ \cite{granda2}.

\section{New \textbf{HDE} and standard Chaplygin gas}
The equation of state of a prefect fluid, standard Chaplygin gas
(SCG), is given by
\begin{eqnarray}\label{Chap}
 p_D=\frac{-A}{\rho_D},
 \end{eqnarray}
 where A is a positive constant, $P_D$ and $\rho_D$ are the pressure and energy
density, respectively. Substituting the equation of state of SCG
(i.e., Eq. \ref{Chap}) into the relativistic energy conservation
equation, leads to  evolving density as
\begin{eqnarray}\label{rhochap}
\rho_D=\sqrt{A+\frac{B}{a^6}}.
\end{eqnarray}
where $B$ is an integration constant. The energy density and
pressure of the scalar field, regarding the SCG dark energy is
written as
\begin{eqnarray}\label{rhophi}
\rho_\phi=\frac{1}{2}\dot{\phi}^2+V(\phi)=\sqrt{A+\frac{B}{a^6}},\\
p_\phi=\frac{1}{2}\dot{\phi}^2-V(\phi)=\frac{-A}{\sqrt{A+\frac{B}{a^6}}},
\label{pphi}
\end{eqnarray}
Hence, it is easy to obtained the scalar potential and the kinetic
energy terms for the SCG model as
\begin{eqnarray}\label{vphi}
&&V(\phi)=\frac{2Aa^6+B}{2a^6\sqrt{A+\frac{B}{a^6}}},\\
&&\dot{\phi}^2=\frac{B}{a^6\sqrt{A+\frac{B}{a^6}}}. \label{ddotphi}
\end{eqnarray}
In this section, first we establish the correspondence between new
HDE and SCG model and re-construct the potential and the dynamics of
scalar fields in non-flat universe, then we discuss the limiting
case of flat universe. Assuming non-flat universe, by equating the
energy density of SCG, Eq.(\ref{rhochap}), and energy density of new
HDE, Eq. (\ref{rhoh}), the constant $B$ can be obtained as
\begin{eqnarray}\label{B}
B=a^6\left(-A+9M_p^4\Big(\frac{\alpha-\beta}{\alpha-\beta-1}ke^{-2x}+\gamma
e^{-\frac{2}{\beta}(\alpha-1)x}\Big)^2\right).
\end{eqnarray}
Using Eqs.(\ref{Chap}), (\ref{wh2}) and (\ref{rhochap}), we have
\begin{eqnarray}
\omega_{\Lambda}=\frac{p_d}{\rho_d}=\frac{-A}{A+Ba^{-6}}=-\frac{1}{3}\left(\frac{k\Big(\frac{\alpha-\beta}{\alpha-\beta-1}\Big)e^{-2x}+\gamma\Big(\frac{3\beta-2\alpha+2}{\beta}\Big)e^{-\frac{2}{\beta}(\alpha-1)x}}
{k\Big(\frac{\alpha-\beta}{\alpha-\beta-1}\Big)e^{-2x}+\gamma
e^{-\frac{2}{\beta}(\alpha-1)x}}\right),\label{wh33}
\end{eqnarray}
Inserting Eq.(\ref{B}) in (\ref{wh33}), one can obtain the constant
$A$ as
\begin{eqnarray}\label{A}
A=9M_p^4\Big(\frac{\alpha-\beta}{\alpha-\beta-1}ke^{-2x}+\frac{3\beta-2\alpha+1}{\beta}\gamma
e^{-\frac{2}{\beta}(\alpha-1)x}\Big)\Big(\frac{\alpha-\beta}{\alpha-\beta-1}ke^{-2x}+\gamma
e^{-\frac{2}{\beta}(\alpha-1)x}\Big)
\end{eqnarray}
Substituting $A$ and $B$ in Eqs.(\ref{vphi}) and (\ref{ddotphi}), we
can rewrite the scalar potential term as
\begin{eqnarray}\label{Vphi}
V(\phi)=M_p^2\left(\frac{2(\alpha-\beta)}{\alpha-\beta-1}ke^{-2x}+\frac{3\beta-\alpha+1}{\beta}\gamma
e^{-\frac{2}{\beta}(\alpha-1)x}\right)
\end{eqnarray}
 and kinetic energy term as
\begin{eqnarray}\label{dotphi4}
\dot{\phi}=M_p\sqrt{\left(\frac{2(\alpha-\beta)}{\alpha-\beta-1}ke^{-2x}+\frac{2(\alpha-1)}{\beta}\gamma
e^{-\frac{2}{\beta}(\alpha-1)x}\right)}
\end{eqnarray}
Using $\dot{\phi}=\phi^{\prime}H$, where prime denotes the
derivative with respect to $x=\ln{a}$, we have
\begin{eqnarray}\label{phiprime}
\phi^{\prime}=M_p\sqrt{\frac{\left(\frac{2(\alpha-\beta)}{\alpha-\beta-1}ke^{-2x}+\frac{2(\alpha-1)}{\beta}\gamma
e^{-\frac{2}{\beta}(\alpha-1)x}\right)}{\frac{1}{\alpha-\beta-1}ke^{-2x}+\gamma
e^{-\frac{2}{\beta}(\alpha-1)x}}}
\end{eqnarray}
Integrating Eq.(\ref{phiprime}), one can obtain the evolutionary
treatment of scalar field as
\begin{eqnarray}\label{phint}
\phi(a)-\phi(0)=M_p
\int_{0}^{x}\sqrt{\frac{\left(\frac{2(\alpha-\beta)}{\alpha-\beta-1}ke^{-2x}+\frac{2(\alpha-1)}{\beta}\gamma
e^{-\frac{2}{\beta}(\alpha-1)x}\right)}{\frac{1}{\alpha-\beta-1}ke^{-2x}+\gamma
e^{-\frac{2}{\beta}(\alpha-1)x}}}dx
\end{eqnarray}
where we take $\ln{a_0}=0$ for present time. Here we established the
connection between new HDE and SCG models and reconstructed the
potential and the dynamics of new HDE model to describe the non-flat SCG cosmology.\\
In the limiting case of flat universe, putting $k=0$ in
Eqs.(\ref{B}) and (\ref{A}), the constants $A$ and $B$ of SCG, are
reduced as
\begin{eqnarray}\label{Bflat}
B=a^6 \Big(-A+9M_p^4 \gamma^2 e^{-\frac{4}{\beta}(\alpha-1)x}\Big)
\end{eqnarray}
\begin{equation}
A=9M_p^4 \frac{3\beta-2\alpha+1}{\beta}\gamma^2
e^{-\frac{4}{\beta}(\alpha-1)x}
\end{equation}
The scalar potential and kinetic energy terms are reduced to
\begin{equation}\label{f1}
V(\phi)=M_p^2\frac{3\beta-\alpha+1}{\beta}\gamma
e^{-\frac{2x(\alpha-1)}{\beta}}
\end{equation}

\begin{equation}\label{f2}
\dot{\phi}=M_p\sqrt{\frac{2(\alpha-1)}{\beta}\gamma
e^{-\frac{2}{\beta}(\alpha-1)x}}
\end{equation}
Using $\dot{\phi}=\phi^{\prime}H$, where prime denotes the
derivative with respect to $x=\ln{a}$, we have
\begin{equation}\label{phiprime2}
\phi^{\prime}=M_p\sqrt{\frac{2(\alpha-1)}{\beta}}
\end{equation}
The evolutionary form of scalar field can be obtained by integrating
of Eq.(\ref{phiprime2}) as follows
\begin{equation}\label{evo_form}
\phi(a)-\phi(0)=M_p\int_0^x\sqrt{\frac{2(\alpha-1)}{\beta}}dx=M_p\sqrt{\frac{2(\alpha-1)}{\beta}}\ln{a}
\end{equation}
\section{new \textbf{HDE} and generalized chaplygin gas}
Although, the SCG model can interpret the accelerated expansion of
universe, but it does not solve the cosmological problems like
structure formation and cosmological perturbation power spectrum
\cite{sanvik}. Subsequently, the SCG was modified to the following
form \cite{Ben}
\begin{equation}\label{gen-rho}
p_D=\frac{-A}{\rho_D^\eta},
\end{equation}
called generalized Chaplygin gas (GCG). Two free parameters involved
in GCG: one is $A$ and the other $\eta$.  Similar with SCG, The GCG
fluid behaves like dust for small size of the universe while it acts
as cosmological constant when universe gets sufficiently large.
Using the energy conservation equation for GCG, the evolving energy
density is obtained as
\begin{equation}\label{gen-rho1}
\rho_D=(A+Ba^{-3\delta})^\frac{1}{\delta}
\end{equation}
where $\delta=\eta+1$. Regarding the scalar field model, the energy
density and pressure of GCG is given by
\begin{eqnarray}\label{rhophi}
\rho_\phi=\frac{1}{2}\dot{\phi}^2+V(\phi)=(A+Ba^{-3\delta})^{-\delta},\\
p_\phi=\frac{1}{2}\dot{\phi}^2-V(\phi)=-A(A+B^{-3\delta})^{-\frac{\delta-1}{\delta}},
\label{pphi}
\end{eqnarray}
Hence, the scalar potential and kinetic energy term can be obtained
as
\begin{eqnarray}\label{vphi-gen}
&&V(\phi)=\frac{2A+Ba^{-3\delta}}{2(Ba^{-3\delta})^{\frac{\delta-1}{\delta}}},\\
&&\dot{\phi}^2=\frac{Ba^{-3\delta}}{(A+Ba^{-3\delta})^{\frac{\delta-1}{\delta}}}\label{dotphi-gen}.
\end{eqnarray}
In this section we establish the connection between new HDE and GCG
models and reconstruct the scalar field DE model. First we assume a
general non-flat universe, then we discuss the limiting case of flat
universe.
In non-flat case, by equating Eq.(\ref{rhoh}) with (\ref{gen-rho1}),
the constant $B$ can be obtained as
\begin{eqnarray}\label{B-gen}
B=a^{3\delta}\left(-A+(3M_p^2)^{\delta}\Big(\frac{\alpha-\beta}{\alpha-\beta-1}ke^{-2x}+\gamma
e^{-\frac{2}{\beta}(\alpha-1)x}\Big)^{\delta}\right)
\end{eqnarray}
 Using Eqs.(\ref{gen-rho}),
(\ref{gen-rho1}) and (\ref{wh2}), we have
\begin{eqnarray}\label{omega_gen3}
\omega_{\Lambda}=\frac{p_d}{\rho_d}=\frac{-A}{(A+Ba^{-3\delta})}=-\frac{1}{3}\left(\frac{k\Big(\frac{\alpha-\beta}{\alpha-\beta-1}\Big)e^{-2x}+\gamma\Big(\frac{3\beta-2\alpha+2}{\beta}\Big)e^{-\frac{2}{\beta}(\alpha-1)x}}
{k\Big(\frac{\alpha-\beta}{\alpha-\beta-1}\Big)e^{-2x}+\gamma
e^{-\frac{2}{\beta}(\alpha-1)x}}\right),\label{wh3}
\end{eqnarray}
Inserting  Eq.(\ref{B-gen}) in (\ref{omega_gen3}), the constant $A$
is obtained as
\begin{equation}\label{A-gen}
A=(3M_p^2)^{\delta}\Big(\frac{\alpha-\beta}{\alpha-\beta-1}ke^{-2x}+\gamma
e^{-\frac{2}{\beta}(\alpha-1)x}\Big)^{\delta-1}\Big(\frac{\alpha-\beta}{\alpha-\beta-1}ke^{-2x}+\frac{3\beta-2\alpha+2}{\alpha-\beta-1}\gamma
e^{-\frac{2}{\beta}(\alpha-1)x}\Big)
\end{equation}
Substituting Eqs.(\ref{B-gen}) and (\ref{A-gen}) in
Eqs.(\ref{vphi-gen}) and (\ref{dotphi-gen}), using Eq.(\ref{H2-1}),
the potential and dynamics of new HDE generalized Chaplygin gas is
obtained as follows
\begin{eqnarray}\label{Vphig}
V(\phi)=M_p^2\left(\frac{2(\alpha-\beta)}{\alpha-\beta-1}ke^{-2x}+\frac{3\beta-\alpha+1}{\beta}\gamma
e^{-\frac{2}{\beta}(\alpha-1)x}\right)
\end{eqnarray}
\begin{eqnarray}\label{dotphi4g}
\dot{\phi}=M_p\sqrt{\left(\frac{2(\alpha-\beta)}{\alpha-\beta-1}ke^{-2x}+\frac{2(\alpha-1)}{\beta}\gamma
e^{-\frac{2}{\beta}(\alpha-1)x}\right)}
\end{eqnarray}
which are exactly same as Eqs.(\ref{Vphi}) and (\ref{dotphi4}) for
new HDE-SCG model. Therefore, the potential and the dynamics of new
HDE-GCG coincide with  new HDE-SCG model. Following the same steps
as done for the SCG model, the evolutionary form of scalar field
describing the new HDE-GCG is as follows
\begin{eqnarray}\label{phintg}
\phi(a)-\phi(0)=M_p
\int_{0}^{x}\sqrt{\frac{\left(\frac{2(\alpha-\beta)}{\alpha-\beta-1}ke^{-2x}+\frac{2(\alpha-1)}{\beta}\gamma
e^{-\frac{2}{\beta}(\alpha-1)x}\right)}{\frac{1}{\alpha-\beta-1}ke^{-2x}+\gamma
e^{-\frac{2}{\beta}(\alpha-1)x}}}dx
\end{eqnarray}
 The potential and the dynamics of new HDE model which describe the GCG cosmology is similar with SCG cosmology.\\
In the limiting case of flat universe, the Hubble parameter and EoS
parameter of new HDE are given by Eqs. (\ref{H2-flat}) and
(\ref{wflat}). Using Eq.(\ref{H2-flat}) and putting $k=0$ in Eqs.
(\ref{B-gen})
 and (\ref{A-gen}), the constants $B$ and $A$ for GCG in flat universe are reduced as follows
 \begin{equation}
 B=a^{3\delta}\Big(-A+(3M_p^2\gamma
 e^{-\frac{2}{\beta}(\alpha-1)x})^{\delta}\Big)
\end{equation}
\begin{equation}
A=\frac{3\beta-2\alpha+2}{\beta}\Big(3M_p^2\gamma
e^{-\frac{2}{\beta}(\alpha-1)x}\Big)^{\delta}
\end{equation}

Same as non-flat case, the scalar potential and the kinetic energy
terms of new HDE, describing the flat GCG cosmology, which can be
obtained by putting $k=0$ in Eqs. (\ref{Vphig}) and
(\ref{dotphi4g}), are same as Eq. (\ref{f1}) and (\ref{f2}) for new
HDE-SCG in flat universe. Therefore, the evolutionary form of scalar
field describing the new HDE-GCG in flat universe is given by Eq.
(\ref{evo_form}).

\section{new \textbf{HDE} and modified  chaplygin gas}
After the GCG was introduced, the new model of Chaplygin gas which
is called modified Chaplygin gas (MCG) was proposed \cite{mcg}.
The equation of state of MCG is written as
\begin{equation}\label{mcg}
p_D=B\rho_D-\frac{A_0}{\rho_D^{\eta}}
\end{equation}
where $B$ and $A_0$ are constant parameters and $0\leq\eta\leq1$. An
interesting feature of MCG is that it can show the radiation era in
the early universe. At the late time, the MCG behaves as
cosmological constant and can be fitted to $\Lambda$CDM model.
 The equation of state and evolving energy density of MCG is given by
\begin{eqnarray}\label{p_mvcg}
p_D=B\rho_D-\frac{A_0a}{\rho_D^{\eta}}&=&\frac{1}{2}\dot{\phi}^{2}-V(\phi
),
\end{eqnarray}
\begin{eqnarray}\label{ro_mvcg}
\rho_D=\Big[\frac{3\delta A_{0}}{[3\delta(B+1)]}%
-\frac{C}{a^{3(\delta)(B+1)}}\Big]^{\frac{1}{\delta}}&=&\frac{1}{2}\dot{\phi}^{2}+V(\phi
),
\end{eqnarray}
where $\delta=\eta+1$, $a$ is a scale factor and $B$, $A_0$, $C$ are constants.\\
From Eqs. (\ref{p_mvcg}) and (\ref{ro_mvcg}), the kinetic and
potential terms of MCG can be obtained as
\begin{eqnarray}\label{dorphi9}
\dot{\phi}^{2} &=&(1+B)\Big[\frac{3\delta A_{0}}{[3\delta(B+1)]}-\frac{C}{a^{3\delta(B+1)}}\Big]%
^{\frac{1}{\delta}}  \nonumber \\
&&-\frac{A_{0}}{\Big[\frac{3\delta A_{0}}{[3\delta(B+1) ]}-\frac{C}{a^{3\delta(B+1)}}\Big]%
^{\frac{\delta-1 }{\delta}}}.
\end{eqnarray}%

\begin{eqnarray}\label{vphi9}
V(\phi ) &=&\frac{(1-B)}{2}\Big[\frac{3\delta A_{0}}{[3\delta(B+1)]}%
-\frac{C}{a^{3\delta(B+1)}}\Big]^{\frac{1}{\delta}}
\nonumber \\
&&+\frac{A_{0}}{2\Big[\frac{3\delta A_{0}}{[3\delta(B+1) ]}-\frac{C}{a^{3\delta(B+1)}}\Big]%
^{\frac{\delta-1 }{\delta}}}.
\end{eqnarray}%
We now establish the correspondence between new HDE and MCG model
and reconstruct the potential and the dynamics of the scalar field
in the presence of new HDE.\\
In non-flat case, by equating Eqs. (\ref{ro_mvcg}) and (\ref{rhoh}),
we get
\begin{equation}\label{C}
C=a^{3\delta(B+1)}\left(\frac{3\delta
A_{0}}{[3\delta(B+1)]}-\Big[3M_{P}^{2}(\frac{\alpha-\beta}{\alpha-\beta-1}ke^{-2x}+\gamma
e^{-\frac{2}{\beta}(\alpha-1)x})\Big]^{\delta}\right).
\end{equation}%
 Using Eqs. (\ref{p_mvcg}) and
(\ref{ro_mvcg}), the EoS parameter of MCG is
\begin{equation}\label{w_dmvcg}
w_{D}=\frac{p_D}{\rho_D}=B-\frac{A_0}{\rho_D^{\eta+1}}
\end{equation}%
Equating $w_D=w_{\Lambda}$ and replacing the energy densities of new
HDE and MCG, $\rho_D=\rho_{\Lambda}$ in Eq.(\ref{w_dmvcg}), we have
\begin{equation}
w_{\Lambda}=B-\frac{A_0}{(3M_{P}^{2}[\alpha
H^{2}+\beta\dot{H}])^{\delta}}=-\frac{1}{3}\left(\frac{k\Big(\frac{\alpha-\beta}{\alpha-\beta-1}\Big)e^{-2x}+\gamma\Big(\frac{3\beta-2\alpha+2}{\beta}\Big)e^{-\frac{2}{\beta}(\alpha-1)x}}
{k\Big(\frac{\alpha-\beta}{\alpha-\beta-1}\Big)e^{-2x}+\gamma
e^{-\frac{2}{\beta}(\alpha-1)x}}\right)
\end{equation}
Therefore, the above equation obtains the value of the parameter
$B_0$ as
\begin{equation}\label{B_0}
A_0=\left(3M_{P}^{2}(\frac{\alpha-\beta}{\alpha-\beta-1}ke^{-2x}+\gamma
e^{-\frac{2}{\beta}(\alpha-1)x})\right)^{\delta}\left(B+\frac{1}{3}\frac{k\Big(\frac{\alpha-\beta}{\alpha-\beta-1}\Big)e^{-2x}+\gamma\Big(\frac{3\beta-2\alpha+2}{\beta}\Big)e^{-\frac{2}{\beta}(\alpha-1)x}}
{k\Big(\frac{\alpha-\beta}{\alpha-\beta-1}\Big)e^{-2x}+\gamma
e^{-\frac{2}{\beta}(\alpha-1)x}}\right)
\end{equation}
Substituting Eqs. (\ref{C}) and (\ref{B_0}) in Eqs. (\ref{dorphi9})
and (\ref{vphi9}), we can re-write the scalar potential and kinetic
energy terms as
\begin{eqnarray}\label{Vphivg}
V(\phi)=M_p^2\left(\frac{2(\alpha-\beta)}{\alpha-\beta-1}ke^{-2x}+\frac{3\beta-\alpha+1}{\beta}\gamma
e^{-\frac{2}{\beta}(\alpha-1)x}\right)
\end{eqnarray}
\begin{eqnarray}\label{dotphi4vg}
\dot{\phi}=M_p\sqrt{\left(\frac{2(\alpha-\beta)}{\alpha-\beta-1}ke^{-2x}+\frac{2(\alpha-1)}{\beta}\gamma
e^{-\frac{2}{\beta}(\alpha-1)x}\right)}
\end{eqnarray}
It is intersecting to note that the above potential and kinetic
energy expressions for the new HDE-MCG model coincide with those
obtained for new HDE-SCG and new HDE-GCG models. Hence, like
previous models, the evolutionary form of scalar field describing
the MCG cosmology is given by
\begin{eqnarray}\label{phintg}
\phi(a)-\phi(0)=M_p
\int_{0}^{x}\sqrt{\frac{\left(\frac{2(\alpha-\beta)}{\alpha-\beta-1}ke^{-2x}+\frac{2(\alpha-1)}{\beta}\gamma
e^{-\frac{2}{\beta}(\alpha-1)x}\right)}{\frac{1}{\alpha-\beta-1}ke^{-2x}+\gamma
e^{-\frac{2}{\beta}(\alpha-1)x}}}dx
\end{eqnarray}
In flat case, same as previous sections, the Hubble parameter and
EoS parameter for new HDE model is given by Eqs. (\ref{H2-flat}) and
(\ref{wflat}) respectively. Putting $k=0$ in Eqs. (\ref{C}) and
(\ref{B_0}), the constants $C$ and $B_0$ for MCG are reduced as
\begin{equation}\label{Cf}
C=a^{3\delta(B+1)}\left(\frac{3\delta
A_{0}}{[3\delta(B+1)]}-\Big[3M_{P}^{2}\gamma
e^{-\frac{2}{\beta}(\alpha-1)x}\Big]^{\delta}\right).
\end{equation}%
\begin{equation}\label{B_0f}
A_0=\left(3M_{P}^{2}\gamma
e^{-\frac{2}{\beta}(\alpha-1)x}\right)^{\delta}\left(B+\frac{3\beta-2\alpha+2}{3\beta}\right)
\end{equation}
Subsequently, the potential and kinetic energy terms are reduced as
to Eq. (\ref{f1}) and (\ref{f2}). Eventually, same as new HDE-SCG
and new HDE-GCG models, the evolutionary form of scalar field
describing the new HDE-MCG model in flat universe is given by Eq.
(\ref{evo_form}).

\section{Squared speed for new HDE and SCG, GCG, MCG models} One of
the crucial physical quantities in the theory of linear perturbation
is the squared speed of sound, $v^2$. The sign of $v^2$ is important
for determining the stability or instability of a given perturbed
mode. The positive sign (real value of speed) indicates the periodic
propagating mode for a density perturbation and in this case we
encounter with the stability for a given mode. The negative sign
(imaginary value of speed) shows an exponentially growing mode for a
density perturbation, means the instability for a given mode
\cite{myung}.\\ Generally, the evolution of sound speed in the
linear regime
 of perturbation is dependent on the dynamics of background cosmology as
 follows \cite{beca}
 \begin{equation}
 v^2=\frac{dp}{d\rho}=\frac{1}{3H}\frac{d}{dH}\Big[H^2(q-\frac{1}{2})\Big],
 \end{equation}
 where $q$ is the deceleration parameter. Therefore, the sign of
 $v^2$ is linked to the sign of $q$ and to the transition epoch from
 CDM dominated phase to DE dominated phase. From the above
 description, one can easily find that the growth of perturbation in
 linear theory is dependent on the choice of DE model
 of background dynamics.\\
 In this section, we calculate $v^2$  for new HDE model as well as
for SCG, GCG and MCG models. The squared speed is introduced as
\begin{equation}\label{sound1}
v^2=\frac{dp}{d\rho}=\frac{\dot{p}}{\dot{\rho}}
\end{equation}
 For SCG model, using Eq.(\ref{Chap}) and
$p_D=w_D\rho_D$, we have
\begin{equation}
v^2=\frac{A}{\rho^2}=-w_{D}
\end{equation}
Assuming the SCG as a quintessence DE model with $-1<w_D<0$, the
squared
speed is positive. Thus, the SCG is stable against density perturbation at any cosmic scale factor.\\
In the case of GCG model, by using Eq.(\ref{gen-rho}), $v^2$ is
obtained as
\begin{equation}
v^2=-\eta w_D
\end{equation}
In the range of quintessence model $-1<w_D<0$ for GCG model, we see
that the GCG model is instable against the density perturbation
($v^2<0$) if $\eta<0$ and stable ($v^2>0$) if $\eta>0$. In MCG
model, by using Eq.(\ref{mcg}), and considering the range of EoS of
MCG as $-1<w_D<0$, $v^2$ can be obtained as
\begin{equation}
v^2=-w_D+2B=|w_D|+2B
\end{equation}
 Therefore, the sing of $v^2$ is positive (stability) if
$B\geq0$. For $B<0$, the sign of sound speed can be positive and
also can be negative. In this case, $v^2$ is positive (stability) if
$|w_D|<2|B|$ and it is negative (instability) if $|w_D|>2|B|$.\\
We now calculate the squared sound speed for new HDE model and
compare it with SCG, GCG and MCG models. First, we assume a general
non-flat case. Differentiating the equation of state, $p=w\rho$ with
respect to time, we have
\begin{equation}\label{newhol9}
\dot{p}=\dot{w_{\Lambda}}\rho+w_{\Lambda}\dot{\rho}
\end{equation}
Inserting Eq. (\ref{newhol9}) in right hand side of
Eq.(\ref{sound1}), and using Eq.(\ref{contin}), the squared sound
speed for new HDE model is obtained as follows
\begin{equation}\label{squar2}
v^2=w_{\Lambda}-\frac{\dot{w_{\Lambda}}}{3H(1+w_{\Lambda})}
\end{equation}
Inserting Eq.(\ref{wh1}) in (\ref{squar2}) and using
Eq.(\ref{H2-1}), the squared sound speed in non-flat universe is
obtained as
\begin{equation}\label{square3}
v^2=-\frac{1}{3}\frac{(\alpha-1)(-2\alpha+3\beta+2)H^2+ke^{-2x}[\beta(\beta-1)+2\alpha-2]}{\beta[(\alpha-1)H^2+ke^{-2x}(\beta-1)]}
\end{equation}
Let us consider the special case $k=1$, the squared sound speed at
present time is reduced as follows:
\begin{equation}\label{square4}
v^2=-\frac{1}{3}\frac{(\alpha-1)(-2\alpha+3\beta+2)(\frac{1}{\alpha-\beta-1}+\gamma)+[\beta(\beta-1)+2\alpha-2]}{\beta[\frac{\alpha-1}{\alpha-\beta-1}+\gamma(\alpha-1)+(\beta-1)]}
\end{equation}
Adopting the best fit values: $\alpha=0.8824$ and $\beta=0.5016$,
obtained by Y. Wang and L. Xu in non-flat universe \cite{wang3}, one
can easily see that $v^2$ in Eq.(\ref{square4}) is positive for
$\gamma<-0.7560$ and negative when $\gamma>-0.7560$. Therefore, at
the present time, new HDE model is stable provided that$\gamma<-0.7560$ and instable when $\gamma>-0.7560$.\\
Here we discuss the validity of the correspondence between new HDE
and different Chaplygin gas models which are driven in previous
sections. For $\gamma<-0.7560$, new HDE is stable and on the other
hand the SCG, GCG with $\eta>0$ and also MCG with $B\geq0$ or
($B<0,|w_D|>2|B|$) are also stable against density perturbation.
Therefore, in this case, a correspondence between new HDE with SCG,
GCG and MCG is not problematic and the reconstructed potential and
dynamics of scalar field for new HDE model which describe the SCG,
GCG and MCG cosmology are logical. However, in this case, one cannot
encounter with the exponentially growing mode of the perturbations.
The new HDE is instable for $\gamma>-0.7560$, Also the GCG model
with $\eta<0$ and MCG with ($B<0, |w_D|<2|B|$) are also instable.
Hence, the correspondence between new HDE with GCG and MCG models
are logical. In this case, the correspondence between new HDE and
SCG is problematic, since the SCG has a stability against
perturbation. In this case, the potential and the dynamics of scalar
field constructed from the evolutionary form of new HDE cannot describe the SCG cosmology.\\
In the limiting case of flat universe, by putting $k=0$ in
Eq.(\ref{square3}), the squared sound speed for new HDE model is
obtained as
\begin{equation}\label{square5}
v^2=-1+\frac{2}{3}\frac{\alpha-1}{\beta},
\end{equation}
which is same as Eq.(\ref{wflat}). Therefore, in flat case, we have
$v^2=w_{\Lambda}$. Adopting the best fit values: $\alpha=0.8502$ and
$\beta=0.4817$, obtained by Y. Wang and L. Xu in flat universe
\cite{wang3}, We can see that $v^2$ in Eq.(\ref{square5}) is
obtained as $-1.21$. Hence, the new HDE model in flat universe is
instable at any scale factor. In flat universe, the correspondence
between new HDE and GCG with $\eta<0$ and MCG with
($B<0,|w_D|<2|B|$) are logical. While the correspondence between new
HDE and SCG, GCG with $\eta>0$ and MCG with ($B\geq0$ or
$B<0,|w_D|>2|B|$) is problematic. From the viewpoint of linear
perturbation theory, the reconstructed potential and dynamics of
scalar field for new HDE model can describe the Chaplygin cosmology,
if both of DE models have a same sign of squared sound speed.
Otherwise, the correspondence between them is problematic.

\section{Conclusion}
In the context of new HDE model, the event horizon IR cut-off is
replaced by new IR cut-off containing the local quantities of Hubble
and time derivative Hubble scales \cite{garanda}. The new HDE model
not only gives the accelerated expansion of the universe, but also
avoids the causality problem appearing with event horizon IR
cut-off. On the other hand, among the several candidates for DE, we
consider the Chaplygin gas model which unifies DE and dark matter.
In this work, we established a correspondence between the new HDE
density and various models of Chaplygin gas scalar field models of
DE in non-flat universe. The non-flatness of universe with small
positive curvature $\Omega_k \sim0.02$ is favored by recent
experimental data \cite{Sie,Uzan}. We adopted the viewpoint that the
scalar field models of DE are effective theories of an underlying
theory of DE. Thus, we should be capable of using the scalar field
models to mimic the evolving behavior of the new HDE and
reconstructing these scalar field models. We reconstructed the
potential and the dynamics of scalar field for new HDE model which
describe the SCG, GCG and MCG cosmology.
 In the limiting case of flat universe, we obtained the simple
analytical solution for the evolutionary form of the SCG, GCG and
MCG scalar fields. We also concluded that, according to the
evolutionary behavior of the new HDE model, the potential and the
dynamics of MCG scalar field are same as the potential and the
dynamics of GCG and MCG models. Finally, from the viewpoint of
linear perturbation theory, we studied the correspondence between
new HDE with SCG, GCG and MCG models. We showed that the
reconstructed potential and dynamics of scalar field for new HDE
model can describe the Chaplygin cosmology, if both the new HDE and
Chaplygin gas DE models have a same sign of squared sound
speed.\\

\textbf{Acknowledgements}\\
We would like to thank K. Karami for reading the manuscript and
giving useful comments.

\end{document}